\documentclass{appolb}
\usepackage{graphicx}
\usepackage{slashed}
\pdfoutput=1

\begin{document}

\title{Probing Space-Time Noncommutativity in the Bhabha Scattering
}
\author{Linda Ghegal
\address{Mentouri Constantine1 University, Route Ain El Bey 25017, Constantine, Algeria.}
}
\maketitle
\begin{abstract}
We investigate Bhabha scattering with the Seiberg-Witten expended noncommutative standard model scenario to first order of the noncommutativity parameter $\Theta_{\mu \nu }$. This study is based on the definition of the noncommutativity parameter that we have assumed. We explore the noncommutative scale $\Lambda_{_{NC}}\geq0.8$ TeV considering different machine energy ranging from $0.5$ TeV to $1.5$ TeV. 
\end{abstract}
  
\section{Introduction and motivation}
The noncommutative standard model based on the noncommutativity of the space and time variables. The noncommutative space-time is deformation of the ordinary one that can be realized by representing ordinary space-time coordinates ${x}^{\mu}$ by Hermitian operators $\hat{x}^{\nu}$ 
\begin{equation}
   \left[\hat{x}^{\mu},\hat{x}^{\nu}\right] =i\Theta^{\mu \nu} =i\frac{c^{\mu \nu }}{\Lambda
_{NC}^{2}}
\end{equation}
Where ${c}^{\mu \nu}$ are dimensionless parameters and $\Lambda_{NC}$ is the energy scale where the noncommutative effects of the space-time will be relevant.
In the present work, we are especially interested in the spacetime noncommutative standard model based on the Moyal-Weyl (WM) product \cite{Ref1} and Seiberg-Witten (SW) Maps \cite{Ref2}.
In the WM product formalism, to define a field theory on noncommutative spacetime, we replace the ordinary ordinary product by the WM $\star$-product 
\begin{eqnarray}
(f\star g)(x)
& = & \exp \left( \frac{1}{2}\Theta _{\mu \nu }\partial _{x^{\mu
}}\partial _{y^{\nu }}\right) f(x)g(y)\mid _{y=x}
\end{eqnarray}
and the SW maps, express noncommutative fields and parameters as local
functions of the commutative fields and parameters. 
\\
The existence of the Seiberg-Witten map to all orders, leads to expansions for matter field $\psi$, gauge fields $V_{\alpha }$ and gauge parameters $\lambda_{\alpha }$ as following 
\begin{equation}
\hat{\psi}=\hat{\psi}\left[ \psi ,V\right] =\psi +\frac{1}{2}\Theta ^{\mu
\nu }V_{\mu }\partial _{\nu }\psi +\frac{i}{8}\Theta ^{\mu \nu }\left[
V_{\mu },V_{\nu }\right] \psi +O\left( \Theta ^{2}\right)
\end{equation}
\begin{equation}
\hat{V}_{\alpha }=\hat{V}_{\alpha }\left[ V\right] =V_{\alpha }+\frac{1}{4}
\Theta ^{\mu \nu }\{\partial _{\mu }V_{\alpha }+F_{\mu \alpha },V_{\nu
}\}+O\left( \Theta ^{2}\right)
\end{equation}
\begin{equation}
\hat{\lambda}_{\alpha }=\hat{\lambda}_{\alpha }\left[ \lambda ,V\right]
=\lambda _{\alpha }+\frac{1}{4}\Theta ^{\mu \nu }\{\partial _{\mu }\lambda
_{\alpha },V_{\nu }\}+O\left( \Theta ^{2}\right)
\end{equation} 

The full description of the minimal noncommutative standard model (mNCSM), built on the $U(1)_{Y}\otimes SU(2)_{L}\otimes SU(3)_{C}$ group, and the complete Feynman rules derived from NCSM, including new interactions between the gauge bosons in the non-minimal NCSM version, are presented in \cite{Ref3}. The extension of SM to noncommutative space-time with motivations coming from string theory and quantum gravity provides interesting phenomenological implications since scale of noncommutativity could be as low as a few TeV, which can be explored at the present or future colliders. Due to the breaking of Lorentz invariance for fixed $\Theta_{\mu \nu }$ background, noncommutativity of space time leads to dependence of cross section on azimuthal angle which is absent in Standard Model (SM) as well as in other models beyond the SM. Thus, the azimuthal dependence of the cross section is a typical signature of noncommutativity and can be used in order to discriminate it against other new physics effects. We have found this dependence to be best suited for deriving the sensitivity bounds on the noncommutative scale $\Lambda_{_{NC}}$, and this is the main purpose of this work, is to derive the bounds on the noncommutative scale $\Lambda_{_{NC}}$ in the Bhabha scattering in the context of mNCSM, by using the framework introduced in \cite{Ref3}. 

The present manuscript will be organised as follows. In the next section we will present some bounds for the noncommutativity parameter $\Theta_{\mu \nu }$ provided by the most recent papers. We will then conclude with the results that we have abtained of noncommutative scale $\Lambda_{_{NC}}$ in the  Bhabha scattering, in the framework of the mNCSM, using the SW maps to the first order of the NC parameter $\Theta_{\mu \nu }$. 

\section{Overview of bounds on noncommutative scale}
We will dedicate this section  to the bounds which can presently be studied in several processes and systems, including estimates for some experiments [4-16]. Most of them used noncommutative antisymetric constant matrix $\Theta_{\mu \nu }$ analogous to the eletromagnetic field strength tensor; denote the time-like components ${c}^{oi}$ by $\vec{E}$ and the space-like components ${c}^{ij}$ by $\vec{B}$. However, they ofen use two different parameterizations for the noncommutative parameter $\Theta_{\mu \nu }$. The first one, is considered as elementary constant in nature, its direction is fixed in the specific celestial Cartesian reference, so one should take into account these
rotation effects on $\Theta_{\mu \nu }$ in this frame, before moving towards the
phenomenological investigations. In the second parameterizations, the values of the tow vectors are fixed as $\vec{E}=\frac{1}{\sqrt{3}}(\vec{i}+\vec{j}+\vec{k})$ and $\vec{B}=\frac{1}{\sqrt{3}}(\vec{i}+\vec{j}+\vec{k})$. The phenomenological consequences of the noncommutative space-time have been explored in several studies for giving bounds on nonvomutative parameter. From the  
$e^{+}e^{-}\rightarrow q\overline{q}$ subprocess and by comparing with the ALEPH and OPAL data of LEP in \cite{Ref4} found two constraints on the noncommutative space-time scale, $215$ GeV$\leq\Lambda_{_{NC}}\leq240$ GeV and $270$ GeV$\leq\Lambda_{_{NC}}\leq310$ GeV respectively. In \cite{Ref5}, the authors investigate the TeV scale signatures of NC space-time in the $e^{+}e^{-}\rightarrow \gamma\gamma$, $e^{-}\gamma\rightarrow e^{-}\gamma$ and $\gamma \gamma\rightarrow e^{+}e^{-}$ using polarized beams in the mNCSM. The process $e^{+}e^{-}\rightarrow \gamma\gamma$ at the International Linear Collider (ILC), in the framework of non-minimal NCSM, with anomalous triple gauge boson couplings is studied in \cite{Ref6}; knowing that, all these bounds deal with the first parameterization. And for the constant parameterization case, the bound $\Lambda_{_{NC}}\geq400$ GeV found from The Drell-Yan process $pp\rightarrow \gamma ,Z\rightarrow l^{+}l^{-}$ at the Large Hadron Collider (LHC) in the framework of the non-minimal NCSM \cite{Ref7}. The Higgs boson pair production $e^{+}e^{-}\rightarrow H H$ in the linear collider (LC) gives the range $\Lambda_{_{NC}}=0.5-1.0$ TeV \cite{Ref8}.
The study of the Lorentz violation in the Higgs sector in the NCSM \cite{Ref9}, leads to a bound on the noncommutative parameter as large as $\Lambda_{_{NC}}\sim10^{6}$ TeV. 
The Higgs boson production process $e^{+}e^{-}\rightarrow Z H$ and SM forbidden process $e^{+}e^{-}\rightarrow H H$ were investigated in the framework of the mNCSM \cite{Ref10}, with Feynman rules involving all orders of the noncommutative parameter which are derived using reclusive formation of SW map. And by applying the same Feynman rules, the authors in \cite{Ref11} have studied the Higgs plus Z-boson production at a future electron-positron collider to explore the sensitivity of future accelerator experiments to noncommutativity, and obtained as lower limit of $\Lambda_{_{NC}}=1.1$ TeV. The bound $\Lambda_{_{NC}}\geq0.5$ TeV determined in \cite{Ref12} from the same process at the futur Linear Collider. 
An experimental constraint on $\Lambda_{_{NC}}$ and projected sensitivities for the future collider
with neutrino-electron Scattering was summarized in \cite{Ref13}. The authors in \cite{Ref14} have also explored an experiment to probe the effects of noncommutative structure in quantum optical.

Since the models considered by the various authors are different, and since also the experimental setups probe different energy and length scales, therefore, one must be careful in the comparison of the various results of the noncommutative scale $\Lambda_{_{NC}}$.
  
\section{Results and conclusion}
In this section, we investigate the effect of space-time noncommutativity on Bhabha scattering, with the ansatz for noncommutative parameter $\Theta_{\mu \nu }$ that we have assumed and we provide the numerical results of our investigation. We are restricting ourselves only to the first order in $\Theta$; thus the interference between SM and NC term can provide required corrections to cross section.
In order to check the sensitivity of Bhabha scattering in the context of noncommutative space-time and specially the phenomenological effects resulting, we have defined the noncommutative parameter with the help of the gamma matrices, which the
noncommutative structure is determined by some spinor background on which the
gamma-dependent $\Theta_{\mu \nu }$ acts.
\begin{equation}
c^{\mu \nu }=\frac{1}{2}\left( \sigma ^{\mu \nu }+\left( \sigma ^{\mu \nu
}\right) ^{+}\right)\label{path}
\end{equation} 
knowing that 
\begin{equation}
\sigma ^{\mu \nu }=\frac{i}{2}\left( \gamma ^{\mu }\gamma ^{\nu }-\gamma
^{\nu }\gamma ^{\mu }\right)
\end{equation}
where $\gamma ^{\mu }$ are Dirac matrices.
\\
\\
We study now, how the space-time noncommutativity affects the $e^{-}(p_{1})e^{+}(p_{2})\rightarrow e^{-}(p_{3})e^{+}(p_{4})$  scattering process, through the exchange of $\gamma$ and $Z$ bosons at tree level (Via the $s$ and $t$ channel). The Feynman rules to the first order in $\Theta$ are given in \cite{Ref3}. 
\\
Feynman rule for

$e(p_{in})-e(p_{out})-\gamma (k)$ vertex 
\begin{equation}
=ieQ_{f}\left[ \gamma _{\mu }-\frac{i}{2}k^{\upsilon }\left( \Theta _{\mu
\nu \rho }p_{in}^{\rho }-\Theta _{\mu \nu }m_{f}\right) \right]
\end{equation}

$e(p_{in})-e(p_{out})-Z(k)$ vertex 
\begin{eqnarray}
=\frac{ie}{\sin 2\theta _{W}}\{(\gamma _{\mu }-\frac{i}{2}k^{\nu }\Theta
_{\mu \nu \rho }p_{in}^{\rho })\left( C^{f}_{V}-C^{f}_{A}\gamma _{5}\right) \nonumber \\
-\frac{i}{2}\Theta _{\mu \nu }m_{f}[p_{in}^{\nu }\left(C^{f}_{V}-C^{f}_{A}\gamma _{5}\right) -p_{out}^{\nu }\left(C^{f}_{V}+C^{f}_{A}\gamma _{5}\right) ]\} 
\end{eqnarray}
where $\Theta _{\mu \upsilon \rho }=\Theta _{\mu \nu }\gamma _{\rho }+\Theta
_{\nu \rho }\gamma_{\mu }+\Theta _{\rho \mu }\gamma _{\upsilon }$,
$C^{f}_{V}=T^{f}_{3}-2Q_{f}\sin^{2}\theta _{W}$ and
$C^{f}_{A}=T^{f}_{3}$   
\\
with $\theta _{W}$ is the Weinberg angle and 
$Q_{f}=\mp1$ for $e^{\mp}$. 
\\
also $p_{out}\Theta p_{in}=p_{out}^{\mu }\Theta _{\mu \nu }p_{in}^{\nu }=-p_{in}\Theta p_{out}$. The momentum conservation reads as $p_{in}+k=p_{out}$.
\\

\begin{figure}[htb]
\centerline{%
\includegraphics[width=12.5cm]{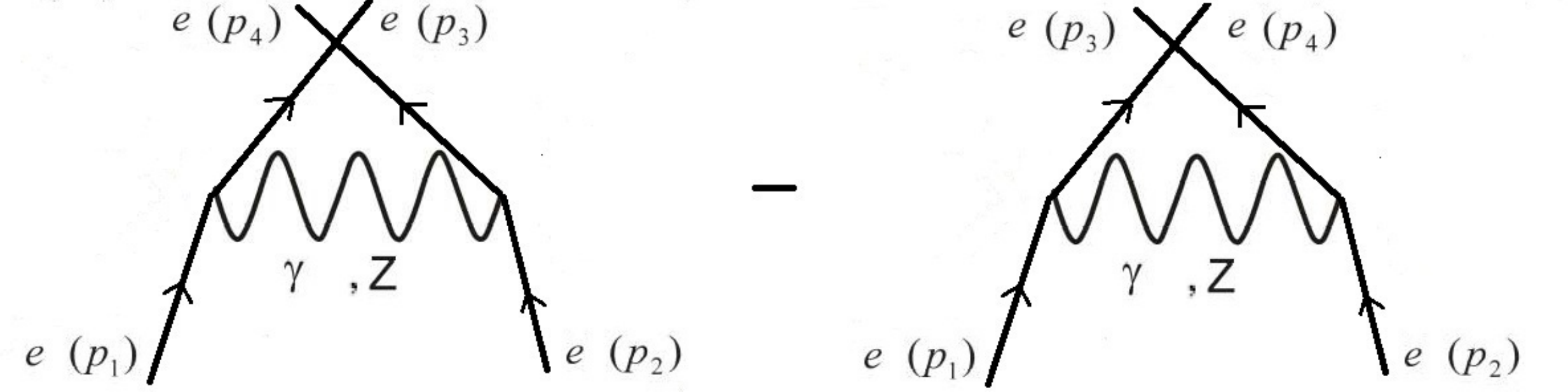}}
\caption{Feynman diagrams for the process $e^{-}e^{+}\rightarrow (\gamma,Z)\rightarrow e^{-}e^{+}$
in the NCSM}
\label{Fig:F1H}
\end{figure}

The corresponding Feynman diagrams are shown in Fig.~\ref{Fig:F1H}
\\
\\
The scattering amplitude to the first order in $\Theta$ can be written as

For the $\gamma$ mediated diagram
\begin{eqnarray}
A^{\gamma}=A^{\gamma}_s+A^{\gamma}_t 
\end{eqnarray}
where

$A^{\gamma}_s=e^{2}\left[1+\frac{i}{2}(p_{2}\Theta p_{1}+p_{4}\Theta p_{3})\right] $
\begin{eqnarray}
\times\left[ \overline{v}(p_{2},s_{2})\gamma^{\mu}u(p_{1},s_{1})\overline{u}(p_{3},s_{3})\gamma_{\mu}v(p_{4},s_{4})\left(\frac{i}{s}\right)\right]   
\end{eqnarray}
and

$A^{\gamma}_t=-e^{2}\left[1-\frac{i}{2}(p_{3}\Theta p_{1}+p_{4}\Theta p_{2})\right]$
\begin{eqnarray}
\times\left[ \overline{u}(p_{3},s_{3})\gamma^{\mu}u(p_{1},s_{1})\overline{v}(p_{2},s_{2})\gamma_{\mu}v(p_{4},s_{4})\left(\frac{i}{t}\right)\right]
\end{eqnarray}

For the $Z$ mediated diagram
\begin{eqnarray}
A^{Z}=A^{Z}_s+A^{Z}_t 
\end{eqnarray}
where

$A^{Z}_s=\frac{e^{2}}{\sin^{2} 2\theta _{W}}
\left[1+\frac{i}{2}(p_{2}\Theta p_{1}+p_{4}\Theta p_{3})\right]$ 
\begin{eqnarray}
\times\left[ \overline{v}(p_{2},s_{2})\gamma^{\mu}\Gamma^{-}_{A}u(p_{1},s_{1})\overline{u}(p_{3},s_{3})\gamma_{\mu}\Gamma^{-}_{A}v(p_{4},s_{4})\left(\frac{i}{s-m^{2}_{Z}}\right)\right]  
\end{eqnarray}
and

$A^{Z}_t=
-\frac{e^{2}}{\sin^{2} 2\theta _{W}}\left[1-\frac{i}{2}(p_{3}\Theta p_{1}+p_{4}\Theta p_{2})\right]$
\begin{eqnarray}
\times\left[\overline{u}(p_{3},s_{3})\gamma^{\mu}\Gamma^{-}_{A}u(p_{1},s_{1})\overline{v}(p_{2},s_{2})\gamma_{\mu}\Gamma^{-}_{A}v(p_{4},s_{4})\left(\frac{i}{t-m^{2}_{Z}}\right)\right]
\end{eqnarray}

where $\Gamma^{\pm}_{A}=\left(C^{e}_{V}{\pm} C^{e}_{A}\gamma _{5}\right)$, 

with $s=\left(p_{1}+ p_{2}\right)^{2}$, $t=\left(p_{1}- p_{3}\right)^{2}$ and $u=\left(p_{1}- p_{4}\right)^{2}$ 
\\

The spin averaged squared-amplitude is given by (See the Appendix \ref{App} for more details).
\begin{equation}
\left\vert \bar{A}\right\vert^{2}=\frac{1}{4}\sum_{spins}\left\vert {A^{\gamma}+A^{Z}}\right\vert^{2} 
\end{equation}

The differential cross section can be written as 
\begin{equation}
\frac{d\sigma }{d\Omega }=\frac{1}{64\pi ^{2}s}\left\vert \bar{A}\right\vert
^{2}
\end{equation}
where $\theta $ and $\phi$ are polar and azimuthal angles respectively,
with $\frac{d\sigma }{d\Omega }=d\cos\theta d\phi$. 

We can obtain the cross section $\sigma =\sigma \left( \sqrt{s},\Lambda
_{NC},\theta ,\phi \right) $ as
\begin{equation}
\sigma =\int_{-1}^{1}d(\cos \theta )\int_{0}^{2\pi }d\phi \frac{d\sigma }{
d\Omega }
\end{equation}

\begin{figure}[htb]
\centerline{
\includegraphics[angle = 90,width=12.5cm]{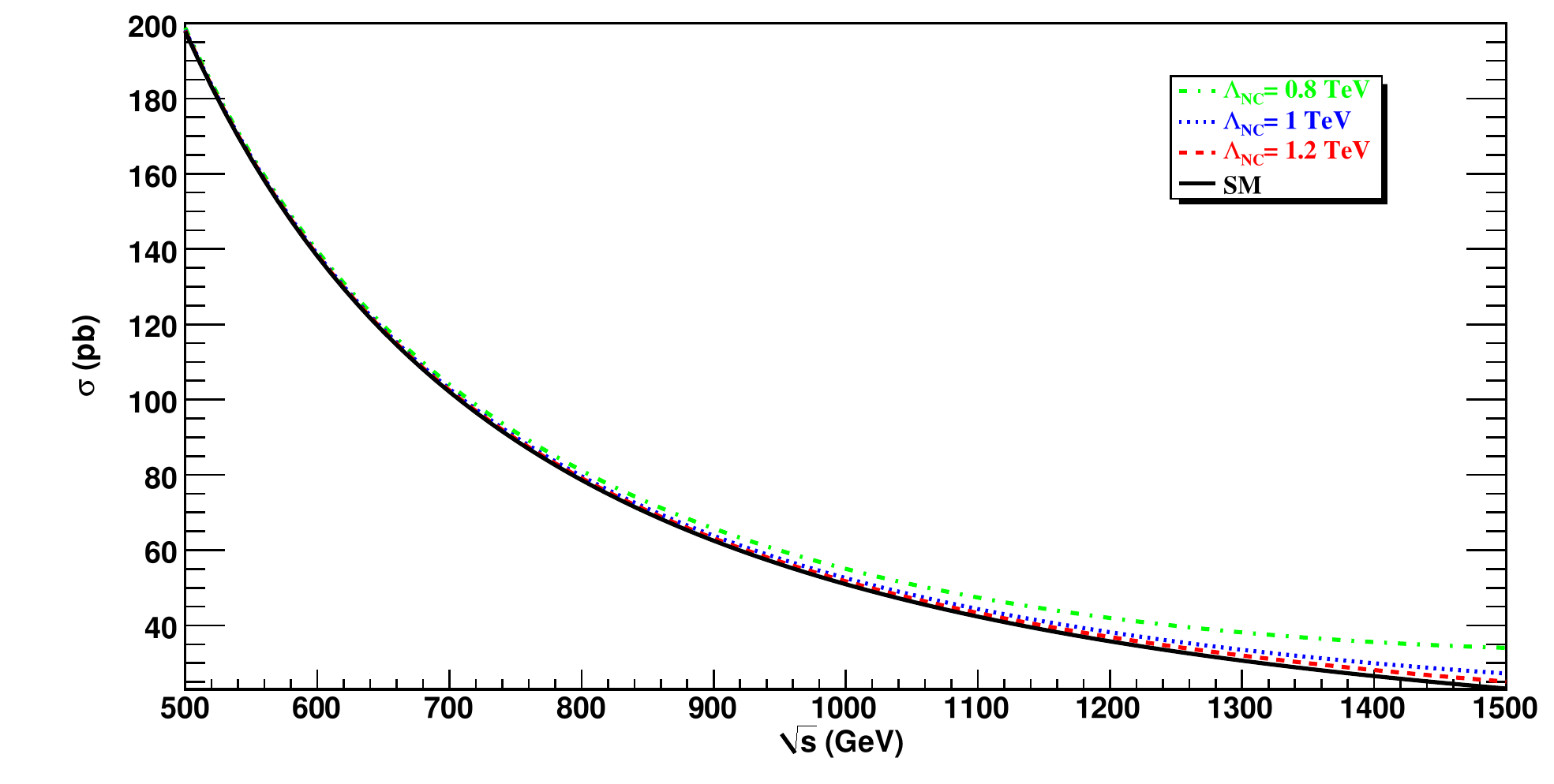}}
\caption{The total cross section for the process $e^{-}e^{+}\rightarrow (\gamma ,Z)\rightarrow e^{-}e^{+}$ [pb] as a function of center of mass energy $E_{com}=\sqrt{s}$ [GeV]}
\label{Fig:F2H}
\end{figure}

Our results are based on Feynman rules for NCSM given in \cite{Ref3}. We analyze the total cross section in the presence of space-time noncommutativity. The results are shown in Fig.~\ref{Fig:F2H} 
\\
The ordinary SM is presented by black curve 
and the NCSM with different curves; green, blue and red, with the corresponding $\Lambda_{_{NC}}=0.8, 1.0$ and $1.2$ TeV, respectively. As can be seen, the noncommutative correction increases on increasing the center of mass energy of the collisions, and we found that the total cross section departs significantly from the standard model value as the machine energy starts getting larger than $1.0$ TeV. Because of the significant sensitivity of the total cross section, it can be used to set lower limits on the noncommutative scale. 

The NCSM is one of the extension for physics beyond SM; Its phenomenological implications are quite interesting since scale of non commutativity could be as low as a few TeV, which can be explored at present or future colliders. In the present work, We have examined the testable nature of noncommutative standard model by analyzing the fundamental process Bhabha scattering at high energy positron-electron linear collider. We have parameterized the noncommutative relationship in terms of noncommutative scale $\Lambda_{_{NC}}$ and anti symmetric matrix ${c}^{\mu \nu}$, which is defined with the help of the gamma matrices, which the noncommutative structure is determined by some spinor background on which the gamma-dependent $\Theta_{\mu \nu }$ acts, and we have shown how the positron-electron scattering process at the tree level is affected by space-time noncommutativity. The NC effects are found to be significant for $\Lambda_{_{NC}}=0.8, 1.0$ and $1.2$ TeV for the center of mass energy $E_{com} = 1.5$ TeV. We got the same results that has been obtained by \cite{Ref16}, i.e. the asymmetry at $\Lambda_{_{NC}}= 0.8$ TeV is greater than that obtained at $\Lambda_{_{NC}}= 1.2$ TeV.

\section*{Acknowledgements}

Linda GHEGAL would like to express her gratitude to Prof. Fedele LIZZI for his useful discussions, during the peroid of her stay at the University of Naples Federico II, Italy.

\appendix

\section{} \label{App}
The spin squared-amplitude can be written as \cite{Ref16}\\

$\overline{|\mathcal{A}|^2}=\overline{|\mathcal{A}^{\gamma}_s|^2}+\overline{|\mathcal{A}^{\gamma}_t|^2}+\overline{|\mathcal{A}^{Z}_s|^2}+\overline{|\mathcal{A}^{Z}_t|^2}-2 \overline{Re(\mathcal{A}^{\gamma}_s \mathcal{A}^{\gamma*}_t)}-2 \overline{Re(\mathcal{A}^{Z}_s \mathcal{A}^{Z*}_t)}$
\begin{equation}+2 \overline{Re(\mathcal{A}^{\gamma}_s \mathcal{A}^{Z*}_s)}-2 \overline{Re(\mathcal{A}^{\gamma}_t \mathcal{A}^{Z*}_s)}+2 \overline{Re(\mathcal{A}^{\gamma}_t \mathcal{A}^{Z*}_t)}
\end{equation}

with the several terms 

$\overline{|\mathcal{A}^{\gamma}_s|^2}=\frac{e^{4}}{4s^{2}}(1+\frac{1}{4}C^{2})$
\begin{equation}
\times Tr[\slashed p_{2}{\gamma}_{\mu}\slashed p_{1}{\gamma}_{\nu}]Tr[\slashed p_{3}{\gamma}^{\mu}\slashed p_{4}{\gamma}^{\nu}],
\end{equation}

$\overline{|\mathcal{A}^{\gamma}_t|^2}=\frac{e^{4}}{4t^{2}}(1+\frac{1}{4}D^{2})$
\begin{equation}
\times Tr[\slashed p_{1}{\gamma}_{\nu}\slashed p_{3}{\gamma}_{\mu}]Tr[\slashed p_{4}{\gamma}^{\nu}\slashed p_{2}{\gamma}^{\mu}],
\end{equation}

$\overline{|\mathcal{A}^{Z}_s|^2}=\frac{e^{4}}{4(\sin 2\theta _{W})^{4}(s-m^{2}_{Z})^{2}}(1+\frac{1}{4}C^{2})$
\begin{equation}
\times Tr[\slashed p_{1}{\gamma}_{\nu}\Gamma^{-}_{A}\slashed p_{2}{\gamma}_{\mu}\Gamma^{-}_{A}]Tr[\slashed p_{4}{\gamma}^{\nu}\Gamma^{-}_{A}\slashed p_{3}{\gamma}^{\mu}\Gamma^{-}_{A}],
\end{equation}

$\overline{|\mathcal{A}^{Z}_t|^2}=\frac{e^{4}}{4(\sin 2\theta _{W})^{4}(t-m^{2}_{Z})^{2}}(1+\frac{1}{4}D^{2})$
\begin{equation}
\times Tr[\slashed p_{1}{\gamma}_{\nu}\Gamma^{-}_{A}\slashed p_{3}{\gamma}_{\mu}\Gamma^{-}_{A}]Tr[\slashed p_{4}{\gamma}^{\nu}\Gamma^{-}_{A}\slashed p_{2}{\gamma}^{\mu}\Gamma^{-}_{A}],
\end{equation}

$-2\overline{Re(\mathcal{A}^{\gamma}_s \mathcal{A}^{\gamma*}_t)}=-\frac{e^{4}}{2st}Re[(1+\frac{i}{2}C)(1+\frac{i}{2}D)$
\begin{equation}
\times Tr[\slashed p_{1}{\gamma}_{\nu}\slashed p_{3}{\gamma}^{\mu}\slashed p_{4}{\gamma}^{\nu}\slashed p_{2}{\gamma}_{\mu}],
\end{equation}

$+2\overline{Re(\mathcal{A}^{\gamma}_s \mathcal{A}^{Z*}_s)}=\frac{e^{4}}{2(\sin 2\theta _{W})^{2}s(t-m^{2}_{Z})}$
\begin{equation}
\times Re\left[ (1+\frac{1}{4}C^{2})Tr[\slashed p_{1}{\gamma}_{\nu}\slashed \Gamma^{-}_{A} \slashed p_{2}{\gamma}_{\mu}] Tr[\slashed p_{4}{\gamma}^{\nu}\Gamma^{-}_{A}\slashed p_{3}{\gamma}^{\mu}]\right],
\end{equation}

$-2\overline{Re(\mathcal{A}^{\gamma}_s \mathcal{A}^{Z*}_t)}=-\frac{e^{4}}{2(\sin 2\theta _{W})^{2}s(t-m^{2}_{Z})}$
\begin{equation}
\times Re\left[(1+\frac{i}{2}C)(1+\frac{i}{2}D)Tr[\slashed p_{1}{\gamma}_{\nu}\Gamma^{-}_{A}\slashed p_{3}{\gamma}_{\mu}\slashed p_{4}{\gamma}^{\nu}\Gamma^{-}_{A}\slashed p_{2}{\gamma}^{\mu}\right],
\end{equation}

$-2\overline{Re(\mathcal{A}^{\gamma}_t \mathcal{A}^{Z*}_s)}=-\frac{e^{4}}{2(\sin 2\theta _{W})^{2}t(s-m^{2}_{Z})}$
\begin{equation}
\times Re\left[(1-\frac{i}{2}C)(1-\frac{i}{2}D)Tr[\slashed p_{1}{\gamma}_{\nu}\Gamma^{-}_{A}\slashed p_{2}{\gamma}^{\mu}\slashed p_{4}{\gamma}^{\nu}\Gamma^{-}_{A}\slashed p_{3}{\gamma}_{\mu}\right],
\end{equation}

$+2\overline{Re(\mathcal{A}^{\gamma}_t \mathcal{A}^{Z*}_t)}=\frac{e^{4}}{2(\sin 2\theta _{W})^{2}u(u-m^{2}_{Z})}$
\begin{equation}
\times Re\left[ (1+\frac{1}{4}D^{2})Tr[\slashed p_{1}{\gamma}_{\nu}\slashed \Gamma^{-}_{A} \slashed p_{4}{\gamma}_{\mu}] Tr[\slashed p_{2}{\gamma}^{\nu}\Gamma^{-}_{A}\slashed p_{3}{\gamma}^{\mu}]\right],
\end{equation}

$-2\overline{Re(\mathcal{A}^{Z}_s \mathcal{A}^{Z*}_t)}=-\frac{e^{4}}{2(\sin 2\theta _{W})^{4}(u-m^{2}_{Z})(t-m^{2}_{Z})}$
\begin{equation}
\times Re\left[(1+\frac{i}{2}C)(1+\frac{i}{2}D)Tr[\slashed p_{1}{\gamma}_{\nu}\Gamma^{-}_{A}\slashed p_{4}{\gamma}_{\mu}\Gamma^{-}_{A}\slashed p_{2}{\gamma}^{\nu}\Gamma^{-}_{A}\slashed p_{3}{\gamma}_{\mu}\Gamma^{-}_{A}\right].
\end{equation}

The factors $C$ and $D$ are given by
 
$C=p_{2}\Theta p_{1}+p_{4}\Theta p_{3}$

$D=p_{3}\Theta p_{1}+p_{4}\Theta p_{2}$\\

The differential cross section is calculated using the center-of-mass frame for the Bhabha $e^{-}(p_{1})e^{+}(p_{2})\rightarrow (\gamma,Z)\rightarrow e^{-}(p_{3})e^{+}(p_{4})$ scattering process, in which the four-momenta of the incoming and outgoing particles are given by
\begin{center}
$p_{1}=p_{e^{-}}=\frac{\sqrt{s}}{2}(1,0,0,1 )=(E_{1},\vec{P_1}),$
\end{center}
\begin{center}
$p_{2}=p_{e^{+}}=\frac{\sqrt{s}}{2}(1,0,0,-1 )=(E_{2},\vec{P_2}),$ 
\end{center}
\begin{center}
$p_{3}=p_{e^{-}}=\frac{\sqrt{s}}{2}(1,\sin{\theta}\cos{\phi},\sin{\theta}\sin{\phi},\cos{\theta})=(E_{3},\vec{P_3}),$
\end{center}
\begin{equation}
p_{4}=p_{e^{+}}=\frac{\sqrt{s}}{2}(1,-\sin{\theta}\cos{\phi},-\sin{\theta}\sin{\phi},-\cos{\theta})=(E_{4},\vec{P_4}), 
\end{equation}
In evaluating the matrix element square, we have ignored the mass of ingoing particles $(m_{e}\simeq0)$, with $ \vec{P_1}+\vec{P_2}= 0 = \vec{P_3} + \vec{P_4} $. 
In the relativist limit $ s \gg 4m^2$, we get $s=(E_1+E_2)^2=4E^2$ (with $E_1=E_2=E$), $u=-\frac{s}{2}(1+\cos\theta)$ and $t=-\frac{s}{2}(1-\cos\theta)$.

In our analysis, we have assumed an ansatz for $\Theta_{\mu,\nu}$ (See Eq. (\ref{path})). The NC antisymmetric tensor $\Theta_{\mu,\nu}$ is defined with the help of the gamma matrices. Using this definition we may write $C$ and $D$ as follows

\begin{equation}
C=p_{2}\Theta p_{1}+p_{4}\Theta p_{3}=\frac{\sqrt{s}}{4\Lambda_{NC}^{2}}[1+\sin{\theta}(\cos{\phi}+\sin{\phi})], 
\end{equation}
\begin{equation}
D=p_{3}\Theta p_{1}+p_{4}\Theta p_{2}=\frac{\sqrt{s}}{4\Lambda_{NC}^{2}}[\sin{\theta}(\cos{\phi}+\sin{\phi})]. 
\end{equation}

\end{document}